\documentclass[floats,amsmath,nofootinbib,superscriptaddress, preprintnumbers, prd]{revtex4}

\usepackage{amssymb}
\usepackage{amsmath,amssymb,graphicx}
\usepackage{epsf}
\usepackage{epstopdf}
\usepackage {amssymb}
\catcode`@=11
\let\savesort=\NAT@sort@cites
\newcommand\nosort[1]{\edef\NAT@cite@list{#1}}
\def\citenosort#1{\let\NAT@sort@cites=\nosort \cite{#1}%
   \let\NAT@sort@cites=\savesort}
\catcode`@=12 

\newcommand{\nc}{\newcommand}
\nc{\ba}{\begin{eqnarray}}
\nc{\ea}{\end{eqnarray}}
\newcommand{\bmx}{\mathbf{x}}
\nc{\oo}{\overline{\delta \Omega}}
\newcommand{\bmk}{\mathbf{k}}

\begin{document}

\title{Effects of Variable Newton Constant During Inflation}

\author{Ali Akbar Abolhasani}
\email{abolhasani-AT-ipm.ir}
\affiliation{School of Physics, Institute for Research in 
Fundamental Sciences (IPM),
P.~O.~Box 19395-5531,
Tehran, Iran}
\author{Hassan Firouzjahi}
\email{firouz-AT-ipm.ir}
\affiliation{School of Astronomy, Institute for Research in 
Fundamental Sciences (IPM),
P.~O.~Box 19395-5531,
Tehran, Iran}
\author{Mahdiyar Noorbala}
\email{noorbala-AT-ipm.ir}
\affiliation{School of Physics, Institute for Research in 
Fundamental Sciences (IPM),
P.~O.~Box 19395-5531,
Tehran, Iran}

\begin{abstract}
In this paper the effects of time-dependent Newton constant $G$ during inflation are studied.
We present the formalism of curvature perturbations in an inflationary system with a time-dependent Newton constant. As an example  we consider a toy model in which $G$ undergoes a sudden change during inflation. By imposing the appropriate matching conditions 
the imprints of this sharp change in $G$ on curvature perturbation power spectrum are studied. We show that if $G$ increases (decreases) during the transition 
the amplitude of curvature perturbations on large scales decreases (increases). In our model with a sudden change in $G$ a continuous sinusoidal modulations on curvature power spectrum is  induced.  However, in a realistic scenario in which the change in $G$ has some finite time scale we expect these sinusoidal modulations to be damped on short scales. The generated features may be used to explain the observed glitches 
on CMB power spectrum. This puts a bound on $\Delta G$ during inflation of roughly the same order as current bounds on $\Delta G$ during the entire observed age of the universe.

\end{abstract}

\preprint{IPM/A-2012/011}

\maketitle

\section{Introduction}

There are strong evidences that Newton gravitational ``constant'' $G$ did not change considerably during most of the history of our universe.  The evidences range from studies of orbits of planets to pulsar timing, evolution of stellar objects, the cosmic microwave background (CMB) and the effect on the abundance of light elements formed during the period of big bang nucleosynthesis.  These studies yield a bound $|\dot{G}/G|\lesssim10^{-(11\hbox{--}13)}\,{\rm yr}^{-1}$ \cite{Uzan:2002vq,Uzan:2010pm}.  This seems to be a strong bound but actually it corresponds to $|\Delta G/G| \lesssim 0.1$ over the entire history of our universe, which makes it less dramatic.  In addition, it should be noted that many other assumptions are hidden in most of these studies (like taking the other `constants' to be constant).  Of course, it is always possible to come up with a complicated theory of variable $G$ that explains all these observations, but least deviations from the standard working model is usually preferred.  Thus the continued lack of evidence for a variable $G$ has been convincing for most people to accept $G=$constant. There is, however, no direct indication from any of the mentioned experiments that $G$ has to be constant at very early times, for example, during inflation.  This is one of our motivations in this paper to pursue implications of the change in Newton constant during the inflationary era.  

Inflation has emerged as the leading paradigm for early universe and structure formation \cite{Guth:1980zm} which is strongly supported by cosmological observations 
\cite{Komatsu:2010fb}. The simplest models of inflation predict almost scale invariant, almost Gaussian and almost adiabatic perturbations on the CMB. 
In this work we would like to examine the effects of a time-dependent Newton constant in inflationary predictions. We present the  formalism for curvature perturbations in inflationary models with a time-dependent $G$. As a particular example, we consider the toy model in which the reduced Planck mass $\Omega \equiv 1/8\pi G$ undergoes a sharp change from $\Omega= \Omega_-$ to  $\Omega= \Omega_+$ in which both $\Omega_\pm$ are constant. 
We put this change in $\Omega$ at an early stage of inflation so the effects of change in Planck mass is within the CMB observational window. As a result one expects to find local features to be imprinted on curvature perturbations from this sudden change in Planck mass. Indeed there has been a lot of interest in the literature to consider the effects of local features in inflation. These features may originate from a  sudden change in slow-roll conditions, sudden change in the inflaton mass, particle creation during inflation, field annihilations during inflation, change in sound speed of perturbations or change in fluids equation of state
\cite{Starobinsky:1992ts, Leach:2001zf, Adams:2001vc, Gong:2005jr, Joy:2007na, Joy:2008qd, Chen:2006xjb, Chen:2008wn, Chen:2011zf, Chen:2011tu, Hotchkiss:2009pj, Arroja:2011yu, Adshead:2011jq, Abolhasani:2012px, Arroja:2012ae, Romano:2008rr, Battefeld:2010rf, Firouzjahi:2010ga, Battefeld:2010vr, Barnaby:2009dd, Barnaby:2010ke, Zarei:2008nr, Biswas:2010si, Silverstein:2008sg, Flauger:2009ab, Flauger:2010ja, Bean:2008na}. The observational motivations behind these models are  to address the glitches in curvature power spectrum on scales associated with $\ell \sim 20-40$.

The simplest implementation of a variable $G$ is a scalar-tensor theory of gravity.  In fact, there are many theoretical models, like string theory, supergravity and theories of extra dimensions, which involve scalar fields non-minimally coupled to a rank-2 tensor.  Thus there is a natural place for theories of variable $G$ within the framework of fundamental physics.  But to maintain a constant $G$, and hence respect the null experiments mentioned above, authors often choose to change variables to another rank-2 tensor which {\emph is} minimally coupled to matter.  In other words, the metric that describes the geometry of space-time is taken to be the one that appears in the Einstein frame.  We will not do so.  Instead we work with a model where the effective gravitational coupling, given by $\Omega_+$,
has no variation during the standard hot big bang cosmology after inflation but experiences a change during the inflationary period. Having this said, if one insists on performing the field redefinition and goes to the Einstein frame with a constant $G$ throughout, then the nature of the physical phenomena is the same in both frames; it is only a matter of interpretation of what one measures. In particular, the observers may use different clocks and rulers to measure physical processes in these two frames, though the final conclusion is the same for both observers.  

The effects of time-dependent Newton constant during inflation via a conformal coupling with  multiple inflation fields to gravity was recently studied in \cite{White:2012ya}, 
see also \cite{Deruelle:2010ht, Makino:1991sg, Tsujikawa:2004my, Koh:2005qp, Koh:2010kg, Gong:2011qe, Kubota:2011re, Kallosh:2010ug, Linde:2011nh, Ferrara:2010yw, Ferrara:2010in, DeFelice:2011jm, Jamil:2009sq}.

The rest of this paper is organized as follows. In Section \ref{background} we present our setup and the background inflation equations. In Section \ref{perturbations} we present the cosmological perturbation theory for general models with a time-dependent Planck mass.
In Section \ref{sudden-change} we consider the special case in which $\Omega$ undergoes a sudden jump. By imposing the appropriate matching conditions we find the curvature perturbations power spectrum which will be used to put some rough bound on $\Delta G/G$ during inflation in this toy model.  Discussions are given in Section \ref{discussions} followed by some technical issues in Appendix A.

\section{Inflation Background}
\label{background}

In this section we present the inflation background. We are interested in a generic model of inflation with a time-dependent Planck mass with the  action  
\begin{equation}
S = \int  d^4x \sqrt{-g} \left(\frac{\Omega}{2}R + {\cal L}_M \right) ,
\end{equation}
where $\Omega (x^\alpha)$ is a positive dimension-2 scalar function of space-time
which is interpreted as the reduced Planck mass related to Newton constant $G$ via $\Omega = 1/ 8 \pi G$. Also $R$ is the Ricci scalar 
and ${\cal L}_M$ represents the matter sector Lagrangian. 

The standard  Einstein general theory of relativity corresponds to the case in which $\Omega = M_P^2$ is constant.  We do not specify the mechanisms which make $\Omega$ dynamical. In principle, $\Omega$ may be a function of other fields in the system. As these fields run toward their minima  a time-dependence in $\Omega$ is induced. For example, this is the situation in simple models of  inflation with the inflaton field $\phi$ in which $\Omega \sim 1-\xi \phi^2$ with $\xi$ being a conformal coupling \cite{Kofman:2007tr}.    

The action contains the generic matter Lagrangian ${\cal L}_M $. 
Our assumption is that the matter sector (the Standard Model fields) is minimally coupled to gravity with metric $g_{\alpha \beta}$.  Of course, we can eliminate the time-dependence of $\Omega$ by a change of metric field
 \ba
 \label{conformal}
 g_{\alpha \beta } \rightarrow   \tilde { g}_{\alpha \beta }=   \frac{\Omega}{M_P^2} g_{\alpha \beta} \, ,
 \ea
 to switch from the ``Jordan frame'' to the Einstein frame  to ensure that $G$ is constant in space and time.  
It is only the question of clocks and rulers one uses in these two frames and how to interpret the measurements from these two sets of clocks and rulers.

We do not concentrate on  a specific inflationary mechanism and the matter sector which supports inflation. All we need is that there is a period in which the universe has positive acceleration, $\ddot a(t) >0$, where $a(t)$ is the scale factor of the FRW universe with the metric
\ba
\label{FRW}
ds^2 = g_{\alpha\beta}dx^\alpha dx^\beta = -dt^2 + a(t)^2 d \bmx^2 \, .
\ea 
In terms of the equation of state parameter $w$, we demand that during inflation $w< -1/3$ followed by a period of standard big bang cosmology with a radiation or matter dominated universe. Also we do not specify the  mechanism in which inflation ends. 
 
With this general picture in mind the modified Einstein equation is  
\begin{equation}
\Omega \, G_{\mu\nu} + (g_{\mu\nu}\nabla^2 - \nabla_\mu\nabla_\nu)\Omega = T_{\mu\nu},
\end{equation}
where $T_{\mu\nu} = \frac{-2}{\sqrt{-g}} \frac{\delta S_M}{\delta g^{\mu\nu}}$ is the standard energy-momentum tensor associated with the matter action.  

A consistency condition for the above equation of motion is obtained by taking its divergence:
\begin{equation}
\label{consistency}
\nabla^\nu T_{\mu\nu} = -\frac{R}{2}  \nabla_\mu \Omega.
\end{equation}
This means that $T^{\mu}_{\nu}$ is not conserved in the usual sense. Recall that the standard argument leading to $\nabla^\nu T_{\mu\nu}=0$ is based on applying the following mathematical fact to $S_M=\int {\cal L}_M$:  Given a diffeomorphism invariant functional $S_M[g,\phi_i]$ of a tensor field $g$ and some other fields $\phi_i$, the divergence of $\delta S_M/\delta g_{\mu\nu}$ vanishes if $\delta S_M/\delta \phi_i=0$ for all $i$.  In our case, however, $\Omega$ depends on the matter fields, so the equations of motion of those fields receive a contribution from $\int\frac{1}{2}\Omega R$ as well as $\int {\cal L}_M$. However, if we define a total energy momentum tensor, $T_{\mu\nu}^{\rm tot}$,
containing the mater fields contribution in $\Omega$, i.e.  
$T_{\mu\nu}^{\rm tot} = \frac{-2}{\sqrt{-g}} \delta  \int \left(\frac{\Omega}{2}R + {\cal L}_M
\right) / \delta g_{\alpha \beta}  $, then $T_{\mu\nu}^{\rm tot}$ is conserved in the usual sense, i.e. 
$\nabla^\nu T_{\mu\nu}^{\rm tot}=0$. 

Consider the isotropic and homogeneous FRW background with the stress energy tensor in the form
\ba
\label{T-eq}
T^\alpha_\beta = P \delta^\alpha_\beta   + (\rho + P) u^{\alpha} u_\beta \, ,
\ea
in which $\rho$ and $P$, respectively,  are the energy density and  pressure and $u^{\alpha}$ is the fluid velocity four-vector normalized as $u^{\alpha} u_{\alpha }=-1$. 
Assuming that $\Omega= \Omega(t)$ at the FRW background level, the background Einstein equations are 
\ba
\label{rho-eq}
\Omega \left( H^2 +\frac{k}{a^2} \right) + \dot{\Omega}H  &&= \frac{\rho}{3} \, , \\
\label{p-eq}
\Omega \left( 3H^2 + 2\dot{H} + \frac{k}{a^2} \right) + \ddot{\Omega} + 2H\dot{\Omega} &&= -P, \\
2\Omega\frac{\ddot{a}}{a} + H\dot{\Omega} + \ddot{\Omega} &&= -\frac{1}{3}(\rho+3P) \, .
\ea
One can check that only two of these equations are independent. Also the consistency equation 
(\ref{consistency}) can be written as 
\ba
\label{conservation}
\dot{\rho} + 3H(\rho+P) = 3\dot{\Omega} \left(2H^2+\dot{H}+\frac{k}{a^2}\right) = \frac{1}{2} \dot{\Omega}R.
\ea
In particular, in the standard situation in which $\dot \Omega=0$, Eq.~(\ref{conservation}) results in the conventional energy conservation equation. 

In order to obtain inflation, we assume that the equation of state parameter $w\equiv P/\rho < -1/3$ during inflation. In practice, we may take $w \simeq -1$, corresponding to a period of slow-roll inflation. We have not specified the mechanism in which inflation ends. It may be as in conventional scenarios of (p)reheating followed by a radiation dominated universe. 

\section{Perturbations}
\label{perturbations}

In this section we study cosmological perturbation theory in our setup with a general function 
 $\Omega(x^\alpha)$.  We use the conventions of Ref.~\cite{Bassett:2005xm}, except that we work with ${\cal R}_c$ (which is the negative of their ${\cal R}$) to describe the comoving curvature perturbations.
The scalar perturbations of the metric are
\begin{equation}
ds^2 = -(1+2A)dt^2 + 2a\partial_i B dx^idt + a^2[(1-2\psi)\delta_{ij} +2\partial_i\partial_jE]dx^idx^j,
\end{equation}
The fluid four-vector, to linear order in perturbation theory, is $u_\mu = (-1-A,v_i)$ in which 
$v_i \equiv \partial_i v$ where $v$ is the velocity potential. 

\subsection{Gauge invariant perturbation equations}

The details of the perturbations of  Einstein equations are given in Appendix \ref{pert-appendix}.  Defining the gauge invariant variables Newton potential $\Phi$ and Bardeen potential $\Psi$
\ba
\Phi = A - \frac{d}{dt}[a^2(\dot{E} - B/a)] \quad , \quad \Psi = \psi + Ha^2(\dot{E} - B/a)  \, , 
\ea
the $(00)$, $(0i)$, $(ii)$ and $(ij) (i\neq j)$ components of perturbed Einstein equations in the gauge invariant form,  respectively, are
\begin{align}
\label{00}
&-\frac{2\Omega }{a^2}\nabla^2\Psi + 6H \Omega (\dot{\Psi} + H \Phi) +3 \dot \Omega \left(  \dot \Psi  + 2 H \Phi \right) 
 + \frac{1}{a^2}\nabla^2 \oo - 3H \left(H \oo  + \dot \oo \right) = -\overline{\delta\rho} \\
\label{0i}
&2 \Omega \dot \Psi + \frac{1}{a^2} \left(a^2 \Omega\right)^. \Phi - a \left( \frac{\oo}{a}
\right)^. = -(\rho+ P) \bar v \\
\label{ij}
&\left( \Psi- \Phi - \frac{\oo}{\Omega} \right)_{, ij} =0 \\
\label{ii}
&2 \Omega\left[(  3 H^2 + 2 \dot H ) \Phi + \ddot \Psi + 3 H \dot \Psi + H \dot \Phi
\right]  +  \dot \Omega ( \dot \Phi + 2 \dot \Psi + 4 H \Phi)   
+2 \ddot \Omega \Phi - \ddot \oo - 2 H \dot \oo - (  3 H^2 + 2 \dot H ) \oo  
= \overline{\delta P}
\end{align}
in which the gauge invariant quantities $\overline{v}, \overline{\delta \rho}$ and $\overline {\delta P}$,  corresponding respectively to perturbations in velocity potential, energy density and pressure, are defined via
\ba
\overline {\delta \rho} \equiv   \delta \rho - a^2 \dot \rho \left( \dot E - \frac{B}{a} \right)  
\quad , \quad  
\overline {\delta P} \equiv   \delta P - a^2 \dot P \left( \dot E - \frac{B}{a} \right)  
\quad , \quad 
\overline v   \equiv v +  a^2 \left( \dot E - \frac{B}{a} \right) \, .
\ea
Similarly, since $\Omega (x^\alpha)$ is an arbitrary function of space-time  
we also defined the gauge invariant perturbation in $\delta \Omega$ via
\ba
\overline {\delta \Omega} \equiv   \delta \Omega - a^2 \dot \Omega \left( \dot E - \frac{B}{a} \right)  \, . 
\ea

Eq.~(\ref{ij}) is very interesting. This indicates that in our setup with a time-dependent Planck mass, the Newton potential $\Phi$ and the Bardeen potential $\Psi$ are not equal and we have gravitational anisotropy
\ba
\label{diff}
\Psi - \Phi = \frac{\oo}{\Omega} \, .
\ea  
In principle, the fact that $\Phi \neq \Psi$ can have interesting observational consequences. 
Observationally, gravitational lensing and integrated Sachs-Wolfe effects are sensitive to
$\Phi -\Psi$ whereas galaxy peculiar velocity measurements are determined by the Newtonian potential $\Phi$  \cite{Jain:2007yk, Afshordi:2008rd, Carroll:2006jn, Hu:2007pj}.

\subsection{Separate Universe Approach}

It is also instructive to look at the separate universe approach \cite{Wands:2000dp, Lyth:2004gb}
which is very suitable for  $\delta N$ formalism in calculating the curvature perturbation power spectrum \cite{Sasaki:1995aw, Sasaki:1998ug, Lyth-Liddle}.   In the separate universe approach  the perturbations inside a horizon-size patch with length $L \sim 1/H$ are homogenized while a long-wavelength mode $k$ with $k > a H$ can vary smoothly over many patches (universes). In this picture the scale factor inside each patch is given by $a(t, \bmx) = a(t) e^{-\psi(t, \bmx)}$ . 

Here we generalize $\delta N$ formalism in our setup to find an equation for the evolution 
of $\zeta$ on super-horizon scales. For this aim, it is very useful to start with the conservation equation (\ref{consistency}) contracted with the unit normal vector  $n^\mu$ perpendicular to the 
surface of constant $t$
\ba
n^\mu = (1- A, -\nabla^i B ) \, ,
\ea
which yields  
\ba
n^\mu T^\nu_{\mu; \nu} = -\frac{R}{2} n^\mu \nabla_\mu \Omega \, .
\ea
After some calculation, this equation can be written as
\ba
\label{pert1}
\dot{\delta \rho} - \frac{1}{2} (\delta \dot \Omega R + \dot \Omega \delta R) = - 3 H (\delta \rho + \delta P) 
+ 3 (\rho+ P) \dot \psi \, ,
\ea
in which $\delta R$, the change in Ricci scalar to linear order in perturbations,  is given by
\ba
\label{delta R}
\delta R = -\frac{2}{a^2} \nabla^2(\Phi-2\Psi) - 6 \ddot \psi -24 H \dot \psi - 6 H \dot A -12 (2 H^2 + \dot H ) A \, .
\ea
To cast Eq.~(\ref{pert1}) into a useful gauge invariant equation for $\dot \zeta$, let us go to the 
uniform density gauge where $\delta \rho=0$ and $\zeta = -\psi$. In this gauge we have $\delta P = \delta P_{\mathrm{nad}}$ in which $\delta P_{\mathrm{nad}}$ indicates the non-adiabatic pressure perturbations. Considering the super-horizon modes $k/a H \rightarrow 0$,
Eq.~(\ref{pert1}) translates to
\ba
\label{pert3}
\frac{\dot \rho}{H}\dot \zeta = 3 H  \delta P_{\rm nad} + 3 \Delta_\Omega 
\ea
in which
\ba
\Delta_\Omega  \equiv
 \dot \Omega \left[  \ddot \Psi  
+ (2 H - \frac{\dot H}{H})\dot \Psi +  H \dot \Phi + 2 (\dot H + 2 H^2) \Phi \right] -  \left( \dot \Omega 
(2 H + \frac{\dot H}{H} ) \right)^. (\zeta + \Psi) - ( \dot H + 2 H^2) \dot \oo \, .
\nonumber\\
\ea
In the limit where $\Omega = M_P^2$ is constant, Eq.~(\ref{pert3}) reduces to the known 
result \cite{Wands:2000dp, Lyth:2004gb} that $\dot \zeta = -H  \delta P_{\rm nad}/(\rho+P)$. However, in our model with a time-dependent gravitational coupling $\Omega$ we have the additional contribution $\Delta_\Omega $. This indicates that $\zeta$ does change on super-horizon scales even in the model with an adiabatic equation of state with $ \delta P_{\rm nad}=0$. As a result, the change in $\Omega$ can be interpreted as a new source of entropy perturbations. Indeed the non-trivial behavior of entropy perturbations in models of multiple field inflation in which $\Omega$ is a functions of theses fields was studied recently 
in \cite{White:2012ya}.

\subsection{Equation for ${\cal R}_c$}
In this section we provide a second order differential equation for the curvature perturbations 
on comoving surface ${\cal R}_c$
\ba
{\cal R}_c = -\psi + H v = -\Psi + H \overline v
\ea
which is very helpful for our analysis in upcoming sections. 

So far our equations were general. Now consider the special case of adiabatic fluid with known equation of state parameter $w(t)=P/\rho$.  Combining Eqs.~(\ref{00}) and (\ref{ii}) we obtain the following single second order differential 
equation for ${\cal R}_c$ (for the details of the derivation see Appendix \ref{pert-appendix})
\ba
\label{R-eq1}
\dot \Theta + \left(H + \frac{\dot \Omega}{\Omega} - \frac{\dot T}{T}
\right) \Theta + c_s^2 k^2 \left( - \frac{\dot T}{T} \delta \Omega_c +  \left( 3 \dot \Omega  - 2 \Omega \frac{\dot T}{T} \right) {\cal R}_{c \bmk}  \right) =0 \, .
\ea 
Here $c_s$ is the sound speed in comoving gauge  defined via 
$\delta P_c = c_s^2 \delta \rho_c$  in which $\delta P_c$ and $\delta \rho_c$ are the perturbations in  pressure  and energy density in comoving gauge  where $v=0$ and 
\ba
\label{Theta}
\Theta &&\equiv  a^2 \left[ \beta_1 \dot {\cal R}_{c \bmk} - \beta_2 \delta \Omega_c - 
\beta_3  \dot {\delta \Omega_c}  
\right] \\
\beta_1 && \equiv T^{-1} \left[ - 2 \Omega \ddot \Omega + 2 H  \Omega \dot \Omega 
- 4 \dot H \Omega^2 + 3 c_s^2 \dot \Omega^2\right] \\
\beta_2 && \equiv  T^{-1} \left[ - H \ddot \Omega - (5 H^2 + 9 H^2 c_s^2 + 3 \dot H) \dot \Omega - H \Omega (8 \dot H + 12 H^2 + 12 H^2 c_s^2)
\right]\\
\beta_3 && \equiv T^{-1} \left [  \ddot \Omega + ( 3c_s^2 -1)H \dot \Omega + 2 \dot H \Omega \right] \\
T && \equiv \dot \Omega + 2 H \Omega \, .
\ea
and $\delta \Omega_c$, is defined as $\delta \Omega$ on comoving surfaces
\ba
\delta \Omega_c \equiv \delta \Omega + \dot \Omega v = \overline {\delta \Omega} + \frac{\dot \Omega}{H} (\Psi +{\cal R}_c) \, .
\ea
One can check that in the limit where $\Omega=  M_P^2$ is constant, then Eq.~(\ref{R-eq1}) reduces to the standard result \cite{Weinberg:2008zzc}
\ba
\label{R-eq0}
\ddot{\cal R}_{c \bmk} + \left( 3 H + \frac{\ddot H}{\dot H} - 2 \frac{\dot H}{H} \right) \dot {\cal R}_{c \bmk}
+ \frac{c_s^2 k^2}{a^2} {\cal R}_{c \bmk} =0 \, .
\ea

\section{ Models with a Sudden Change in Newton Constant}
\label{sudden-change}  

Having presented our perturbation equations  in the previous section
for a general time-dependent Planck mass $\Omega(t)$, here we present a  toy model. 
Consider a model in which $\Omega$ goes under a sudden change at $t= t_c$
separated by two constant values $\Omega_-$ and $\Omega_+$. In this view, inflation has two stages, the first inflationary stage corresponds to time before phase transition, $t < t_c$, with $\Omega= \Omega_-$ and the second stage of inflation corresponds to time $t> t_c$ with
$\Omega= \Omega_+$. In this picture both $\Omega_\pm$ are constant and in particular 
$\Omega_+ = M_P$ in which $M_P$ is the current Planck mass.

In this approximation we can solve the system of equations (\ref{00}--\ref{ii}) in the limits before the phase transition $t< t_c$ with $\Omega = \Omega_-$ and after the phase transition $t>t_c$  with $\Omega = \Omega_+$ and glue the two solutions with the appropriate matching conditions on the background quantities and on ${\cal R}_c$ and $\dot{\cal R}_c$ at the level of perturbations.  

First consider the background case. The background equations (\ref{rho-eq}) and (\ref{p-eq})
near  $t=t_c$ on each side ($t=t_c^-$ and $t= t_c^+$) are 
\ba
\label{back1}
3 \Omega_-  H_-^2 = \rho_-  \quad , \quad
(3 H_-^2 + 2 \dot H_- ) \Omega_- = - p_-
\ea
and
\ba
\label{back2}
3 \Omega_+  H_+^2 = \rho_+  \quad , \quad
(3 H_+^2 + 2 \dot H_+ ) \Omega_+ = - p_+
\ea
in which $\rho_- = \rho(t_c^-)$ and $\rho_+ = \rho(t_c^+)$ and so on.

Now we consider the matching conditions on the background quantities. Our assumption is that 
across $t=t_c$ the geometry is continuous. At the background level this requires $a(t)$ to be continuous. Integrating Eq.~(\ref{rho-eq}) across $t=t_c$ we have $\int_-^+ \rho d t   = 3 \int_-^+ (\Omega H^2 + \dot \Omega H)$. We demand that there is no $\delta$-function type singularity on background energy density, i.e., the discontinuity in $\rho$ is not worse than the step function. This means that 
$\int_-^+ \rho d t  =0$. Furthermore, we expect that $ \int_-^+ \Omega H^2=0$ since neither $H$ nor $\Omega$ are singular. As a result, we conclude that
\ba
\label{back-bc1}
\int_-^+ H \dot \Omega =0 \, .
\ea 
This means that although $\dot \Omega$ has a $\delta$-function type singularity at $t=t_c$
but $H$ goes to zero at $t=t_c$ so rapidly that $H \dot \Omega \rightarrow 0$ at $t=t_c$.
This is a curious result of our assumption. This means that at $t_c$ the universe expansion 
comes to a halt and $\dot a(t) =0$.

The second matching condition at the background level is obtained from integrating Eq.~(\ref{p-eq}) across the transition which yields 
\ba
\label{back-bc2}
\int_-^+ P dt = 2 \left[ H \Omega  \right]_-^+ \, .
\ea
Here we use the notation that $\left[ X  \right]_-^+ = X(t_c^+) - X(t_c^-) $ for the quantity $X$.
Eq.~(\ref{back-bc2}) indicates that $P$ may have a $\delta$-function type singularity at $t=t_c$.

There are different possibilities on $\rho$, $P$ and $H$ such that the matching conditions 
(\ref{back-bc1}) and (\ref{back-bc2}) can be satisfied. One reasonable assumption can be that the intrinsic properties of the fluid driving inflation before and after the transition do not change. This means that the equation of state parameter $w$ and the sound speed $c_s$ remain unchanged in both periods, i.e.,
\ba
w_- = w_+ \equiv w \quad , \quad c_{s\, -} = c_{s \, +}  \equiv c_s\, .
\ea
Using $P_-= w \rho_-$ and $P_+= w \rho_+$ in Eqs.~(\ref{back1}) and (\ref{back2})
yields 
\ba
\label{cond1} 
\frac{\dot H_-}{H_-^2} = \frac{\dot H_+}{H_+^2} \, .
\ea
We can further assume that there is no jump in energy density so that $\rho_- = \rho_+$. As a result, from the continuity of $w$ one also concludes that there is no jump in pressure
and $P_-= P_+$. These assumptions correspond to the case that there is no jump in the thermodynamical properties of the fluid driving inflation. 
 Note that our assumption in taking $\rho$ and $P$ to be constant may be motivated from our staring assumption that it is the  Jordan frame metric $g_{\alpha \beta}$ (and not $ \tilde g_{\alpha \beta}$ as defined in Eq.~(\ref{conformal})) which minimally couples to the SM fields. In this view it is natural to expect the matter sector properties such as $\rho$ and $P$ to be continuous while the geometrical quantities such as $H$ have non-trivial dynamics due to the change 
 in $\Omega$. However,  at this level we do not provide any model in which these conditions are satisfied. In principle, one can cook up models in which these conditions may be satisfied dynamically. We will come back to this question briefly in  Section \ref{discussions}.

Having studied the matching conditions at the background level, we now consider the matching conditions for the perturbations. Physically, we expect both $\Psi$ and ${\cal R}_c$ to be continuous on the surface of transition. Geometrically, this corresponds to the assumption that the intrinsic and extrinsic curvatures on the three-surface at $t=t_c$ are continuous and
\ba
\label{match0}
\left[ {\cal R}_c \right]_-^+ =0\, .
\ea
Intuitively speaking, from the separate universe approach \cite{Wands:2000dp, Lyth:2004gb},
on each patch we have $a(t, \bmx)= a(t) e^{-\psi(t, \bmx)}$. The continuity of the scale factor on each patch at $t=t_c$ requires $\psi$ to be continuous. Translated in a gauge invariant way, it is reasonable to expect $\Psi$ to be continuous. Similarly  one expects the curvature perturbations on comoving surface 
${\cal R}_c = -\psi + H v$ to be continuous.   The other matching conditions has to be imposed on 
$\dot {\cal R}_c$. The details of the matching conditions are left for Appendix 
\ref{pert-appendix} where it is shown that
\ba
\label{match1}
\left[ \frac{\dot H}{H} \dot {\cal R}_c \right]_-^+ =0  \, .
\ea
Using the condition (\ref{cond1}) this leads to
\ba
\label{match2}
 \dot {\cal R}_{c +} = \beta  \dot{\cal R}_{c -} \quad, \quad  \beta \equiv  \sqrt{\frac{\Omega_+}{\Omega_-}} \, .
\ea 
In the limit where there is no change in the gravitational coupling and $\beta=1$, we obtain the expected result that $\dot {\cal R}_{c +} =  \dot {\cal R}_{c -}$. Also note that $\beta$ can be bigger or smaller than unity. 

Having presented the matching conditions, here we present the incoming and outgoing solutions 
for ${\cal R}_{c \bmk} $. For this purpose, let us first solve the background system. Our assumption is that both periods of inflation are driven by a single fluid with constant equation of state parameter $w$ and sound speed $c_s$. To sustain inflation we require $w < -1/3$ but for practical purposes we consider $w \simeq -1$ corresponding to a slow-roll inflation.  Solving the background equations with $\dot \Omega_\pm =0$ for each inflationary stage one obtains
\ba
\rho= \rho_- \left( \frac{a}{a_c}
\right)^{-3 (1+w)}
\ea
which results in
 \ba
a(\eta) = a_c \left( 1+ \frac{{\cal H}_-}{\gamma} (\eta- \eta_c)
\right)^{\gamma} \quad , \quad
{\cal H}(\eta) = {\cal H}_- \left( 1+ \frac{{\cal H}_-}{\gamma} (\eta- \eta_c)
\right)^{-1} \quad \quad \eta < \eta_c
\ea
and 
 \ba
a(\eta) = a_c \left( 1+ \frac{{\cal H}_+}{\gamma} (\eta- \eta_c)
\right)^{\gamma} \quad , \quad
{\cal H}(\eta) = {\cal H}_+ \left( 1+ \frac{{\cal H}_+}{\gamma} (\eta- \eta_c)
\right)^{-1} \,  \quad \quad \eta > \eta_c
\ea
Here $\eta_c$ represents the conformal time when $\Omega$ undergoes sudden changes 
 and  $a_c$ is the value of the scale factor at that time $a_c = a(\eta_c)$. Also 
${\cal H}$ is the Hubble factor measured in conformal time ${\cal H} = a'/a = a H$
and ${\cal H}_\pm$ represents the value of ${\cal H}$ just before and just after the phase transition: ${\cal H}_- = {\cal H}(\eta_c^-)$ and  ${\cal H}_+ = {\cal H}(\eta_c^+)$.

The dimensionless parameter $\gamma$ is defined via
\ba
\gamma \equiv \frac{2}{1+ 3 w} \, .
\ea
We note that during inflation  $\gamma<-1$ and for nearly slow-roll inflation $\gamma \lesssim -1$. 
The solution for ${\cal R}_c$ in each region of inflation is easily obtained by solving 
Eq.~(\ref{R-eq0}). Defining the Sasaki-Mukhanov variable $u= a {\cal R}_c$, Eq.~(\ref{R-eq0})
can be written in the following form
\ba
\label{SM-eq}
u''_\bmk + \left( c_s^2 k^2 - \frac{a''}{a} \right) u_\bmk =0
\ea
in which the prime denotes the derivative with respect to the conformal time.
Now using the background solution for $a(\eta)$ we have
\ba
\frac{a''}{a}\bigg|_{\eta< \eta_c} = \frac{\gamma (\gamma -1)}{\left( \eta - \eta_c + \gamma {\cal H}_-^{-1} \right)^2} \quad      , \quad 
\frac{a''}{a}\bigg|_{\eta> \eta_c} = \frac{\gamma (\gamma -1)}{\left( \eta - \eta_c + \gamma {\cal H}_+^{-1} \right)^2} 
\ea
Plugging this in Eq.~(\ref{SM-eq}) and imposing the Bunch-Davies vacuum for the initial conditions yields
\ba
\label{R1}
{\cal R}_{c1} = C_1 x_1(\eta) ^\nu H^{(1)}_\nu (x_1(\eta))  
\ea
in which $H^{(1)}_\nu (x) $ is the Hankel function of first kind, 
\ba
C_1 = \frac{-1}{2 \Omega_-  a_c} \left( \frac{i \pi c_s e^{i \pi \nu}}{ 3 k (1+w)}
\right)^{1/2} \, x_1(\eta_c)^{-\nu +1/2} \, ,
\ea
\ba
\nu \equiv \frac{1}{2} - \gamma = \frac{3 (w-1)}{2 (3 w +1)}
\ea 
and 
\ba
x_1(\eta) \equiv - c_s k \left (\eta - \eta_c + \gamma {\cal H}_-^{-1} \right) \, .
\ea
For the outgoing solution, the answer in given as a combination of the Hankel function of the first and the second kind
\ba
\label{R2}
{\cal R}_{c2} =  x_2(\eta)^\nu \left[
 C_2 H^{(1)}_\nu \left(x_2(\eta) \right)   +   D_2 H^{(2)}_\nu \left(x_2(\eta) \right) 
 \right] \, ,
\ea
in which $ C_2$ and $D_2$ are constants of integrations and
\ba
\nu \equiv \frac{1}{2} - \gamma = \frac{3 (w-1)}{2 (3 w +1)}
\ea 
and 
\ba
x_2(\eta) \equiv - c_s k \left (\eta - \eta_c + \gamma {\cal H}_+^{-1} \right) \, .
\ea
Note that in the limit where $\Omega$ is continuous, we have $C_2= C_1$ and $D_2=0$. As  we shall see, any non-trivial effect form change in $\Omega $ is encoded in the ratio 
$|\frac{C_2 - D_2}{C_1}|$. Also note that since ${\cal H}_+ \neq {\cal H}_-$, the variable $x$ is not continuous at
$\eta_c$. Specifically, one can check that
\ba
x_1(\eta_c) = \frac{k}{k_c} \quad , \quad
x_2(\eta_c)  = \beta  \frac{k}{k_c}
\ea
in which we have defined
\ba
k_c \equiv -\frac{{\cal H}_-}{c_s\,  \gamma}\, .
\ea
In this view, $k_c$ can be interpreted as the mode which leaves the horizon at the critical time 
$\eta_c$.

Now imposing the matching conditions (\ref{match0}) and (\ref{match2}) yields
\ba
\label{C2}
\frac{C_2}{C_1} &=&  \frac{i\pi x_2(\eta_c)}{4} \beta^{-\nu} \left[   H_{\nu}^{(1)}\left(x_1(\eta_c )\right) H_{\nu-1}^{(2)}\left(x_2(\eta_c) \right) -  \beta  H_{\nu-1}^{(1)}\left(x_1(\eta_c )\right) H_{\nu}^{(2)}\left(x_2(\eta_c) \right) \right]   \\
\label{D2}
\frac{D_2}{C_1} &=&  \frac{i\pi x_2(\eta_c)}{4} \beta^{-\nu} \left[  \beta H_{\nu-1}^{(1)}\left(x_1(\eta_c )\right) H_{\nu}^{(1)}\left(x_2(\eta_c) \right) -   H_{\nu}^{(1)}\left(x_1(\eta_c )\right) H_{\nu-1}^{(1)}\left(x_2(\eta_c) \right) \right]  
\ea

As usual we are interested in curvature perturbations, ${\cal P}_{\cal R}$, 
for modes which become super-horizon 
by the end of inflation when $\eta \rightarrow 0$
\ba
\label{R-powr}
\langle  {\cal R}_{c \bmk} {\cal R}_{c \bmk'}\rangle \equiv (2\pi)^{3} P_{{\cal R}}(k)~\delta^3(\bmk+\bmk')
\quad , \quad 
{\cal P}_{\cal R}\equiv \frac{k^{3}}{2 \pi^{2}}P_{\cal R}(k)\,
\ea

Using the asymptotic forms of the Hankel function we have
\ba
{\cal R}_{c\bmk} (\eta\rightarrow 0)  \simeq \frac{-i\,  2^{\nu}}{\pi} \Gamma(\nu) (C_2 - D_2)
\ea
Correspondingly, the curvature perturbation power spectrum, ${\cal P}_{\cal R}$, at the end of inflation is obtained to be
\ba
{\cal P}_{\cal R} (\eta=0) = T \, {\cal P}_{{\cal R} 1}
\ea
where ${\cal P}_{{\cal R} 1}$ is the incoming curvature perturbation power spectrum and 
$T$ is the transfer function
\ba
T \equiv \left| \frac{C_2 -D_2}{C_1} \right|^2 \, .
\ea
In the slow-roll limit when $\nu \simeq 3/2$, ${\cal P}_{{\cal R} 1}$  is approximately 
given by
\ba
{\cal P}_{{\cal R}1} \simeq \frac{H^2}{12 (1+ w) \pi^2 \Omega_-^2 c_s } \, .
\ea

Note that in the absence of any phase transition in which $C_2=C_1$ and $D_2=0$, we obtain the expected result that $T=1$.  Using the specific values of $C_2$ and $D_2$ as given
in Eq.~(\ref{D2}) yields
\ba
\label{T}
T =  \frac{\pi^2}{4} \beta^{2(1-\nu)} \left(\frac{k}{k_c}\right)^2  \left[
\left( J_{\nu-1} ( \frac{\beta k}{k_c})  J_{\nu} ( \frac{k}{k_c} ) - 
\beta   J_{\nu} ( \frac{\beta k}{k_c})  J_{\nu-1} ( \frac{k}{k_c} )
\right)^2 +  \left( J_{\nu-1} ( \frac{\beta k}{k_c})  Y_{\nu} ( \frac{k}{k_c} ) - 
\beta   J_{\nu} ( \frac{\beta k}{k_c})  Y_{\nu-1} ( \frac{k}{k_c} )
\right)^2  \right]
\ea
One can check that in the limit where $\Omega_-= \Omega_+$ and $\beta=1$ we have $T=1$.

\begin{figure}
\includegraphics[width =  3in ]{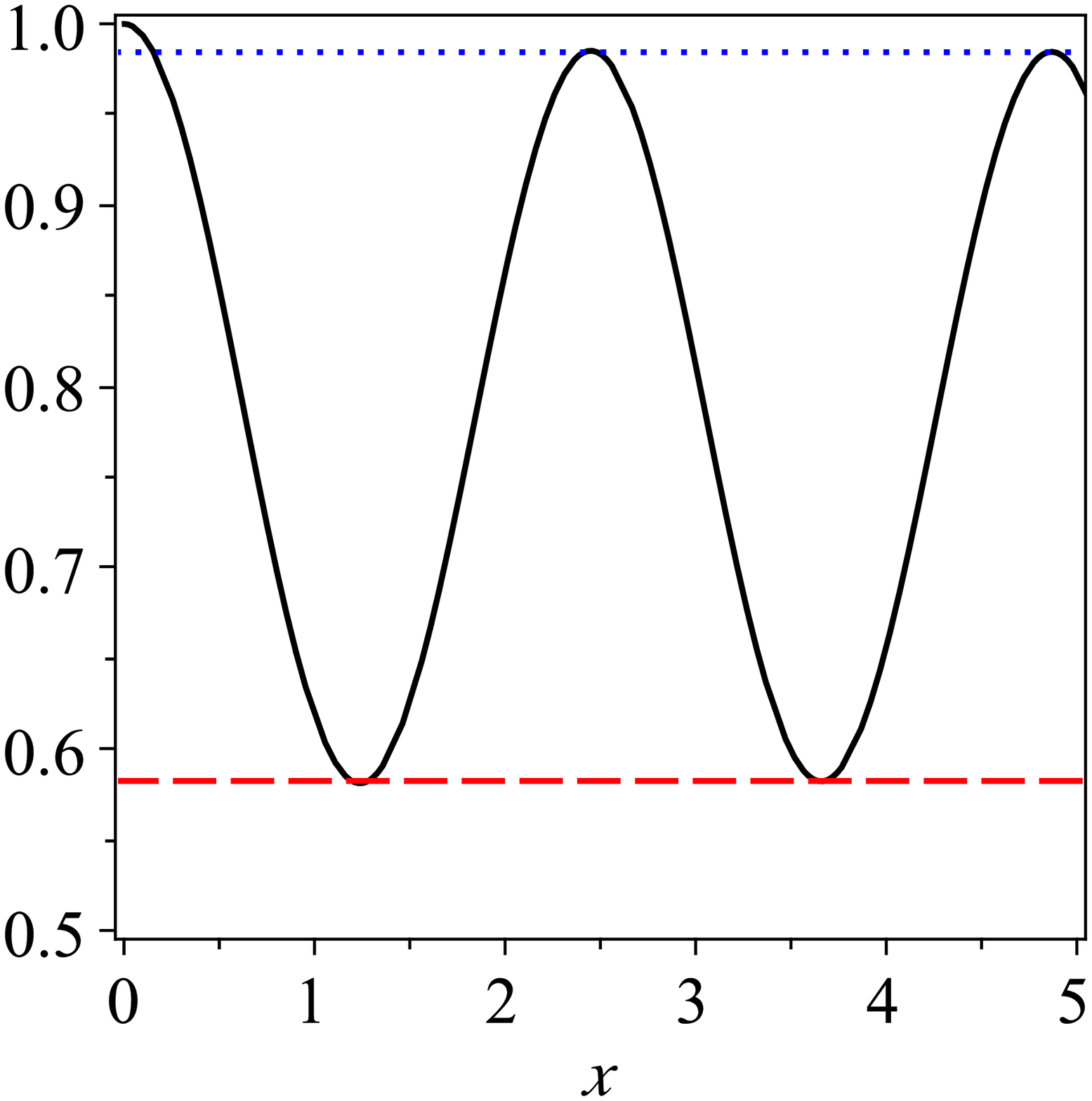}
\hspace{2cm}
\includegraphics[width =  3in ]{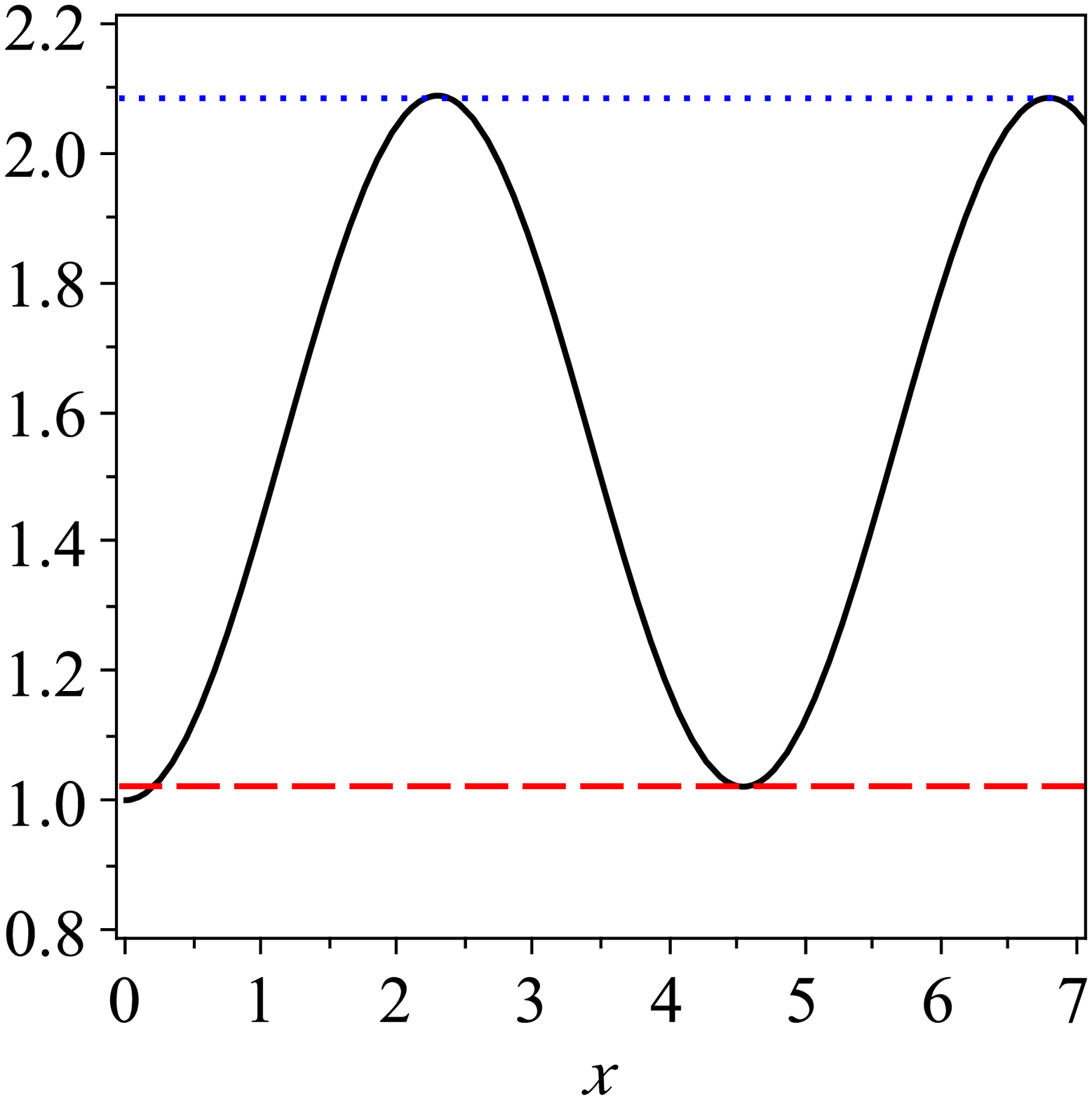}
\caption{
Here we plot the transfer function $T(k)$ in Eq.~(\ref{T}) as a function of $x= k/k_c$.  
The left figure is for $\beta =1.3$ whereas for the right figure $\beta=0.7$, both with 
$\nu=1.53$. The dotted blue line and the dashed red line correspond, respectively, to the maximum and minimum defined in Eqs.~(\ref{beta-great}) and (\ref{beta-great2}).
}
\label{Transfer}
\end{figure}

It is instructive to look at different limits of $T$. Consider modes which left the horizon before 
the time of phase transition, corresponding to  $k \ll k_c$. Using the small argument limit of Bessel functions we obtain 
\ba
T (k\ll k_c) \simeq 1- \frac{\beta^2-1}{2(\nu-1)} x^2 \, ,
\ea
where $x=k/k_c$.   Interestingly, in models in which $\Omega_+ > \Omega_-  (\Omega_- < \Omega_+ ) $  so $\beta>1 (\beta <1 )$, we see a shortage (excess) of power spectrum on large scales.  This is reasonable, since for a given source of energy density, i.e. $\rho_- = \rho_+$, we expect to have less gravitational interaction when the effective Planck mass increases,
i.e., when $\Omega_+ > \Omega_-$. In particular, when $\Omega \rightarrow \infty$, we expect the gravitational decoupling limit in which no metric perturbations are excited.  For a view of 
$T(k)$ see Figure \ref{Transfer}.

Now consider the small scale limit, modes which are deep inside the horizon at the time of phase transition corresponding to $k\gg k_c$. Again, using the large argument limits of Bessel functions one obtains
\ba
T(k \gg k_c) \simeq \beta^{1- 2 \nu} \left[ 1+ (\beta^2 -1) \cos^2 \left(\beta x - \frac{\nu \pi}{2} - \frac{\pi}{4} \right) \right] \, .
\ea
Curiously we note that for modes deep inside the horizon, $k\gg k_c$,  $T$ has a non-decaying sinusoidal modulations as can be  seen in  Figure~\ref{Transfer}. Physically, one may expect that modes which are deep inside the horizon should not be affected by the transition in $\Omega$. 
The non-decaying  sinusoidal modulations on small scales are a result of the assumption that the change in $\Omega$ happens instantaneously. Realistically one expects that the change in $\Omega$ takes some time scale, say $\Delta \eta$. As a result for modes with frequency $\omega > 1/\Delta \eta$, one expects $T$ to be constant with no sin modulation. On the other hand for larger modes with frequency  $\omega \lesssim 1/\Delta \eta$, one expects to see 
localized features in $T$.  In order to verify this prediction, it would be interesting to construct a model in which the change in $\Omega$ has a non-zero time-scale, say one or few e-foldings. 
We would like to come back to this question in a future work.

The spectral index of curvature perturbation power spectrum, $n_s$, is defined via
$n_s -1 \equiv d \ln {\cal P}_{\cal R} /d \ln k$. In the model with a constant Planck mass, $n_s$
is calculated to be
\ba
n_s = 4 - 2 \nu = 1 + \frac{6 (1+ w)}{1+ 3w} \, .
\ea
In order to satisfy the best fit of WMAP observation $n_s =0.96$, we need $w \simeq 1.52$.
Now in our model with a sudden jump in $\Omega$ the power spectrum has sinusoidal modulations and a power law ansatz for the spectral index is not a good one. It is an interesting exercise to check the predictions of our scenario, such as the spectral index, the amplitude of tensor perturbations, etc.\ using the WMAP data analysis, as performed for example in 
\cite{Joy:2008qd, Biswas:2010si},  which predict similar features in power spectrum but in a different context. 

The transfer function at large $k$ behaves differently for $\beta>1$ and $\beta<1$.  In the former case, there is a suppression of power: 
\ba
\label{beta-great}
\beta^{1-2\nu} \leq T   \leq\beta^{3-2\nu}  < 1 \,  \quad  \quad (\beta>1)
\ea
where we have made use of the fact that the observed tilt of the power spectrum is towards red, i.e., $3 - 2\nu = n_s - 1 < 0$.  We may ignore the non-damping oscillations of $T$ at large $k$ with the arguments of the preceding paragraph, but we expect that the first peak/trough in Figure~\ref{Transfer} still survives. This may be employed to address the observed glitches in CMB power spectrum for $\ell \sim 20-40$.  The largest error bar in this region in WMAP data  is not more than one order of magnitude.  Thus let us assume that the suppression of ${\cal P}_{\cal R}$ is not smaller than a factor $1/f$ with $f>1$, e.g., we may take $f=2$.  Then it follows that for $\nu \simeq \frac{3}{2}$ we have $\beta \lesssim \sqrt{f}$.
In terms of Newton constant $G$ we can write
\begin{equation}
  1-f  \lesssim \frac{\Delta G}{G} < 0,
\end{equation}
where $G$ is Newton constant measured today.  This puts an ${\cal O}(1)$ bound on the variation of $G$ during inflation in our model with a sudden  jump in Newton constant. 
Interestingly, this bound is similar to observational constraint on $|\Delta G/G| < 0.1$  during the observed age of the universe as discussed in the Introduction.

In the case $\beta<1$ we have an enhancement of power: 
\ba
\label{beta-great2}
1 < \beta^{3-2\nu} \leq T \leq \beta^{1-2\nu} \quad \quad (\beta <1)  \, .
\ea 
We can now assume that the enhancement of ${\cal P}_{\cal R}$ is smaller than  a factor $f >1$.  Then we find $\beta \gtrsim \frac{1}{\sqrt{f}}$ or equivalently
\begin{equation}
 0<  \frac{\Delta G}{G} \lesssim 1 - \frac{1}{f} \, .
\end{equation}

\section{Discussions}
\label{discussions}

In this paper we considered the effects of change in Newton constant $G$ or 
the reduced Planck mass $\Omega = 1/8 \pi G$ during inflation. The motivations for considering a time-dependent Planck mass may come from models of high energy physics, like string theory, in which the effective four-dimensional Planck mass is a function of extra dimension fields such as the volume modulus. In the standard big bang cosmology there are tight constraints on the variations of $G$. However, there is no direct constraint on the variation of  $G$ during inflation.

In this work  we presented the formalism of cosmological perturbation theory in the models with  a time-dependent $\Omega$. In particular, we have seen that a time-varying $\Omega$ causes  gravitational anisotropy in which the Bardeen potential and the Newton potential are not equal.
This can have interesting implications for the CMB and for lensing. Also, in the context of the separate universe approach, we have seen that a time-dependent $\Omega$ effectively acts as a new source of entropy perturbations. As a result, $\zeta$ does change  on super-horizon scales even for an adiabatic inflationary fluid.  

As a specific solvable example, we considered a toy model in which $\Omega$ undergoes a rapid change from $\Omega_-$ to $\Omega_+$ during inflation. After imposing the appropriate matching conditions on comoving curvature perturbations ${\cal R}_c$ and its derivative $\dot {\cal R}_c$
we obtained the outgoing power spectrum in terms of the incoming power spectrum. The effects of change in $\Omega$ is encoded in the  transfer function $T(k)$. On physical grounds we expect that $T(k)$ should have localized features centered around $k=k_c$, where $k_c$  represents the mode which leaves 
the horizon at the time of phase transition. This is verified in our analysis. In addition, we also see a continuous sinusoidal modulation on $T(k)$. We argued that this is an artifact of our simplifying assumptions that the change in $\Omega$ takes place  instantaneously. In realistic models in which the change in $\Omega$ takes some finite non-zero time one expects that for very small scales $T(k)$ reaches the asymptotic value unity. 

On the other hand, if one chooses to work in the Einstein frame with a constant $G$, then it is natural to ask what mechanisms create these features on power spectrum. As we discussed in Introduction and in Sec.~\ref{background},  in the Einstein frame the matter sector will have a complicated non-standard  Lagrangian.  In this view, in the Einstein frame one expects to see some sudden changes in fluid properties which causes the local features on power spectrum.
In some ways, these sudden changes in fluid or matter sectors may be modeled via any of the mechanisms studied so far in literature, i.e., in \cite{Starobinsky:1992ts, Leach:2001zf, Adams:2001vc, Gong:2005jr, Joy:2007na, Joy:2008qd, Chen:2006xjb, Chen:2008wn, Chen:2011zf, Chen:2011tu, Hotchkiss:2009pj, Arroja:2011yu, Adshead:2011jq, Abolhasani:2012px, Arroja:2012ae, Romano:2008rr, Battefeld:2010rf, Firouzjahi:2010ga, Battefeld:2010vr, Barnaby:2009dd, Barnaby:2010ke, Zarei:2008nr, Biswas:2010si, Silverstein:2008sg, Flauger:2009ab, Flauger:2010ja, Bean:2008na} in the context of Einstein frame gravity.

In our analysis so far, we have not presented a dynamical mechanism which causes the variations in $\Omega$. One interesting realization for creating a dynamical time-dependence 
in $\Omega$ is to couple the gravitational system non-minimally to scalar fields which vary during inflation. In this 
view, $\Omega$ can couple either to the inflaton field or to additional light or heavy fields present in the model. As a  concrete example, consider an inflationary model with two fields: the inflaton field
 $\phi$ and the waterfall field $\chi$ with the matter Lagrangian 
 \begin{equation}
{\cal L}_M = -\frac{1}{2} (\partial\chi)^2 -\frac{1}{2}  (\partial\phi)^2 - V(\chi, \phi).
\end{equation}
Here the potential $V$  is the same as in models of hybrid inflation 
\cite{Linde:1993cn, Copeland:1994vg}
\begin{equation}
V = \frac{1}{2}m^2\phi^2 + \frac{\lambda}{4\lambda}(\chi^2 - \frac{M^2}{\lambda})^2 + \frac{1}{2}g^2\phi^2\chi^2 \, .
\end{equation}
Furthermore, suppose that 
 \ba
 \Omega = \Omega_- + \xi \chi^2 \, ,
 \ea
 in which $\xi$ is a dimensionless number. 
 
The picture we have for the dynamics of the system is similar to waterfall dynamics in hybrid inflation. The system has two distinguished periods, the time before the 
waterfall phase transition corresponding to $\phi < \phi_c= M/g$ and  and the time after the waterfall, $\phi > \phi_c$.  The waterfall field is very heavy during inflation so it is frozen during the first stage of inflation and $\chi=0$. As a result, during the first stage of inflation $\Omega = \Omega_-$. After $\phi$ reaches the critical value $\phi= \phi_c$, the waterfall field becomes tachyonic and rapidly rolls to its global minima $\chi= \pm M/\sqrt \lambda $. 
The quantum fluctuations of the waterfall field  $\delta \chi$ play crucial roles as studied in recent works \cite{Lyth:2010ch, Abolhasani:2010kr, Fonseca:2010nk, Abolhasani:2011yp, Gong:2010zf, Lyth:2012yp, Abolhasani:2012px, Clesse:2010iz, Martin:2011ib, Mulryne:2011ni, Avgoustidis:2011em, Kodama:2011vs, Abolhasani:2010kn, Bugaev:2011qt, Bugaev:2011wy}. 
As a result, at the end of waterfall transition, on each Hubble patch 
$\chi^2 = \langle \delta \chi^2 \rangle  = M^2 /\lambda $ and
\ba
\Omega_+ = \Omega_- + \frac{\xi M^2}{\lambda} \, .
\ea
One can make the waterfall phase transition sharp enough, so the time of change in $\Omega$ is reasonably short, say one e-folding or so. However, it is crucial to note that it is not arbitrarily sharp and in principle it takes a finite non-zero time scale for the waterfall phase transition to complete.  This is expected to eliminate the unwanted non-decaying sin modulations in $T(k)$ for small scales. Also note that, depending on model 
parameters \cite{Abolhasani:2011yp, Abolhasani:2012px}, 
the waterfall transition can happen either  during early stages of inflation or towards the end of inflation. 
Having presented this dynamical mechanism for change in $\Omega$, it is an interesting exercise to study this model in details and see how the waterfall dynamics can be employed to induce a time-dependence in Planck mass. We would like to come back to this question in a future work.

\section*{Acknowledgement}
We would like to thank Eugene Lim, Andrei Linde, 
Mohammad Hossein Namjoo, Misao Sasaki and 
Moslem Zarei  for useful discussions, comments and correspondences.

\appendix
\section{Perturbed Equations and Matching Conditions}
\label{pert-appendix}

Here we provide the perturbed Einstein equations in some details. The $00$, $0i$, $ii$ and $i \neq j$
components of Einstein equations are
\begin{align}
\label{00-Ein}
&- 6 H \Omega (\dot \psi + HA) - 3\dot \Omega \dot \psi  - 6 H \dot \Omega A 
+ 3 H^2 \delta \Omega + 3 H  {\delta \dot \Omega} + \frac{k^2}{a^2}
\left[ \delta \Omega - \dot \Omega \theta - 2 \Omega \psi - 2 H \Omega \theta
\right]  = \delta \rho \\
\label{0i-Ein}
&\dot \psi +( H + \frac{\dot \Omega}{2 \Omega})  A = -\frac{\rho + P}{2 \Omega} v+ 
\frac{\delta \dot \Omega}{2 \Omega}
 - H \frac{\delta \Omega}{2 \Omega} \\
\label{ij-Ein}
&\dot \theta + (H + \frac{\dot \Omega}{\Omega}) \theta = \frac{\delta \Omega}{\Omega} + A -\psi \\
\label{ii-Ein}
&\left[ 2 \Omega ( 2 \dot H + 3 H^2) + 4 H \dot \Omega + 2 \ddot \Omega \right] A
+ (\dot \Omega + 2 H \Omega) \dot A + 2 \Omega \ddot \psi + (2 \dot \Omega + 6 H \Omega) \dot \psi 
- (3 H^2 + 2 \dot H) \delta \Omega - 2 H  \delta \dot \Omega -\delta \ddot \Omega 
 = \delta P
\end{align}
in which $\delta \rho$ and $\delta P$ are the perturbations in energy density and pressure, and 
$v$ is the velocity scalar potential related to the fluid velocity four vector $u_{\mu}$ via
$u_\mu = (-1-A, \partial_i v)$. Furthermore,  we have defined
\ba
\theta \equiv a^2 (\dot E - B/a) \, .
\ea
Playing with these equations one can obtain Eqs.~(\ref{00}-\ref{ii}) in the gauge invariant form. 

To find the matching conditions we can find an equation relating $\dot {\cal R}_c$ to $\Psi$ as follows. To obtain this relation it is very helpful to go to the coming gauge where $v=0$ and ${\cal R}_c =-\psi$.  In this gauge Eqs.~(\ref{00-Ein}),  (\ref{0i-Ein}) and  (\ref{ii-Ein})  can be written as
\ba
\label{00-comoving}
\delta \rho_c =  3 \dot \Omega \dot {\cal R}_c - \frac{3 H}{T} \left[ \dot \Omega \delta \dot \Omega_c - (3 H \dot \Omega + 4 H^2 \Omega) \delta \Omega_c + 2 \Omega \dot \Omega \dot{\cal R}_c \right] + \frac{k^2}{a^2} \left[ \delta \Omega_c + \frac{\dot \Omega}{H} 
(\Psi + {\cal R}_c ) - 2 \Omega \Psi \right] \\
\label{0i-comoving}
-2 \Omega \dot {\cal R}_c + T A = \delta \dot \Omega_c - H \delta \Omega_c &&\\
\label{ii-comoving}
\delta P_c = -6 H \Omega  \dot {\cal R}_c + \frac{F}{T} (  2 \Omega \dot {\cal R}_c - H \delta \Omega_c + \delta \dot \Omega_c ) - 3 (\dot H + H^2) \delta \Omega_c - 3 H \delta \dot \Omega_c
\ea
in which we have defined
\ba
T\equiv \dot \Omega + 2 H \Omega \quad , \quad
F \equiv \ddot \Omega + 2 \dot H \Omega + 2 H \dot \Omega + 6 H^2 \Omega \, .
\ea

Now using the definition $\delta P_c = c_s^2 \delta \rho_c$ we obtain 
\ba
\label{R-dot-eq}
\left[ \frac{2 \Omega F }{T} - 6 H \Omega - 3 c_s^2 \dot \Omega + 
\frac{6 H c_s^2}{T} \Omega \dot \Omega  \right] \dot {\cal R}_c = 
c_s^2 \frac {k^2}{a^2} \left[ \delta \Omega_c + \frac{\dot \Omega}{H} {\cal R}_c - (2 \Omega - \frac{\dot \Omega}{H}) \Psi \right]\nonumber\\
+ \left[ \frac{F H}{T} + 3 \dot H + 3 H^2 + \frac{3 H c_s^2}{T} (3 H \dot \Omega + 4 H^2 \Omega) \right] \delta \Omega_c
+ \left[ -\frac{F}{T} + 3 H - \frac{3 H c_s^2}{T} \dot \Omega   \right] \delta \dot \Omega_c
\ea

Now we look into matching conditions. We consider the model with the sudden change in $\Omega(t)$ from $\Omega_-$ to $\Omega_+$ at $t=t_c$ as presented in Sec.~\ref{sudden-change}.  In this model $\dot \Omega=0$ before and after the transition but $\dot \Omega$
has a $\delta$-function type singularity at $t=t_c$, i.e. $\int_-^+ \dot \Omega = \Omega_+ - \Omega_-$. To simplify  the analysis further, we also assume that $\delta \Omega_c =0$ in both regions which is a reasonable assumption when $\dot \Omega_\pm =0$.  As argued in the main text, the first matching condition
is the continuity of ${\cal R}_c$ across the time of transition 
\ba
\label{R-bc}
\left[ {\cal R}_c \right]_-^+ =0 \, .
\ea
To find the matching conditions on $\dot {\cal R}_c$ we use Eq.~(\ref{R-dot-eq}). Assuming that 
$\Psi$ and $c_s$ are continuous as argued in the main text, and $\dot \Omega_\pm =\delta \Omega_c=0$,  from Eq.~(\ref{R-dot-eq}) we obtain
\ba
\left[ \frac{\dot H}{H} \dot {\cal R}_c  \right]_-^+ =0 \, .
\ea
Now assuming that the fluid equation of state $w$ remains unchanged as argued in the main text, and using Eq.~(\ref{cond1}) we end up with
\ba
\label{dot-R-bc}
\dot{\cal R}_{c+} = \sqrt{\frac{\Omega_+}{\Omega_-}} \dot{\cal R}_{c-} \, .
\ea
The two matching conditions (\ref{R-bc}) and (\ref{dot-R-bc}) are our two matching conditions to express the outgoing solution of ${\cal R}_c$ in terms of the incoming solution. 



\bigskip
\section*{References}
\begin{thebibliography}{}


\bibitem{Uzan:2010pm} 
  J.~-P.~Uzan,
  ``Varying Constants, Gravitation and Cosmology,''
  Living Rev.\ Rel.\  {\bf 14}, 2 (2011)
  [arXiv:1009.5514 [astro-ph.CO]].


\bibitem{Uzan:2002vq} 
  J.~-P.~Uzan,
  ``The Fundamental constants and their variation: Observational status and theoretical motivations,''
  Rev.\ Mod.\ Phys.\  {\bf 75}, 403 (2003)
  [hep-ph/0205340].

\bibitem{Guth:1980zm}
A.~H.~Guth,
``The Inflationary Universe: A Possible Solution To The Horizon
 And Flatness Problems,''
Phys.\ Rev.\ D {\bf 23}, 347 (1981);\\
 K.~Sato,
  ``First Order Phase Transition of a Vacuum and Expansion of the Universe,''
  Mon.\ Not.\ Roy.\ Astron.\ Soc.\  {\bf 195}, 467-479 (1981). \\
A.~D.~Linde,
``A New Inflationary Universe Scenario: A Possible Solution Of The Horizon,
 Flatness, Homogeneity, Isotropy And Primordial Monopole Problems,''
Phys.\ Lett.\ B {\bf 108}, 389 (1982); \\
A.~Albrecht and P.~J.~Steinhardt,
 ``Cosmology For Grand Unified Theories With Radiatively Induced Symmetry
 Breaking,''
Phys.\ Rev.\ Lett.\  {\bf 48}, 1220 (1982).


\bibitem{Komatsu:2010fb}
  E.~Komatsu {\it et al.},
  ``Seven-Year Wilkinson Microwave Anisotropy Probe (WMAP) Observations:
  Cosmological Interpretation,''
  arXiv:1001.4538 [astro-ph.CO].


\bibitem{Starobinsky:1992ts} 
  A.~A.~Starobinsky,
  ``Spectrum of adiabatic perturbations in the universe when there are singularities in the inflation potential,''
  JETP Lett.\  {\bf 55}, 489 (1992)
  [Pisma Zh.\ Eksp.\ Teor.\ Fiz.\  {\bf 55}, 477 (1992)].

\bibitem{Leach:2001zf} 
  S.~M.~Leach, M.~Sasaki, D.~Wands and A.~R.~Liddle,
  ``Enhancement of superhorizon scale inflationary curvature perturbations,''
  Phys.\ Rev.\ D {\bf 64}, 023512 (2001)
  [astro-ph/0101406].

\bibitem{Adams:2001vc}
  J.~A.~Adams, B.~Cresswell, R.~Easther,
  ``Inflationary perturbations from a potential with a step,''
  Phys.\ Rev.\  {\bf D64}, 123514 (2001).
  [astro-ph/0102236].


\bibitem{Gong:2005jr} 
  J.~-O.~Gong,
  ``Breaking scale invariance from a singular inflaton potential,''
  JCAP {\bf 0507}, 015 (2005)
  [astro-ph/0504383].

\bibitem{Joy:2007na}
  M.~Joy, V.~Sahni, A.~A.~Starobinsky,
  ``A New Universal Local Feature in the Inflationary Perturbation Spectrum,''
  Phys.\ Rev.\  {\bf D77}, 023514 (2008).
  [arXiv:0711.1585 [astro-ph]].
  
 \bibitem{Joy:2008qd}
  M.~Joy, A.~Shafieloo, V.~Sahni and A.~A.~Starobinsky,
  ``Is a step in the primordial spectral index favored by CMB data ?,''
  JCAP {\bf 0906}, 028 (2009)
  [arXiv:0807.3334 [astro-ph]].


\bibitem{Chen:2006xjb} 
  X.~Chen, R.~Easther and E.~A.~Lim,
  ``Large Non-Gaussianities in Single Field Inflation,''
  JCAP {\bf 0706}, 023 (2007)
  [astro-ph/0611645].

\bibitem{Chen:2008wn} 
  X.~Chen, R.~Easther and E.~A.~Lim,
  ``Generation and Characterization of Large Non-Gaussianities in Single Field Inflation,''
  JCAP {\bf 0804}, 010 (2008)
  [arXiv:0801.3295 [astro-ph]].

\bibitem{Chen:2011zf} 
  X.~Chen,
  ``Primordial Features as Evidence for Inflation,''
  JCAP {\bf 1201}, 038 (2012)
  [arXiv:1104.1323 [hep-th]].

\bibitem{Chen:2011tu} 
  X.~Chen,
  ``Fingerprints of Primordial Universe Paradigms as Features in Density Perturbations,''
  Phys.\ Lett.\ B {\bf 706}, 111 (2011)
  [arXiv:1106.1635 [astro-ph.CO]].

\bibitem{Hotchkiss:2009pj} 
  S.~Hotchkiss and S.~Sarkar,
  ``Non-Gaussianity from violation of slow-roll in multiple inflation,''
  JCAP {\bf 1005}, 024 (2010)
  [arXiv:0910.3373 [astro-ph.CO]].

\bibitem{Arroja:2011yu} 
  F.~Arroja, A.~E.~Romano and M.~Sasaki,
  ``Large and strong scale dependent bispectrum in single field inflation from a sharp feature in the mass,''
  Phys.\ Rev.\ D {\bf 84}, 123503 (2011)
  [arXiv:1106.5384 [astro-ph.CO]].

\bibitem{Adshead:2011jq} 
  P.~Adshead, C.~Dvorkin, W.~Hu and E.~A.~Lim,
  ``Non-Gaussianity from Step Features in the Inflationary Potential,''
  Phys.\ Rev.\ D {\bf 85}, 023531 (2012)
  [arXiv:1110.3050 [astro-ph.CO]].

\bibitem{Abolhasani:2012px} 
  A.~A.~Abolhasani, H.~Firouzjahi, S.~Khosravi and M.~Sasaki,
  ``Local Features with Large Spiky non-Gaussianities during Inflation,''
  arXiv:1204.3722 [astro-ph.CO].
  
\bibitem{Arroja:2012ae} 
  F.~Arroja and M.~Sasaki,
  ``Strong scale dependent bispectrum in the Starobinsky model of inflation,''
  arXiv:1204.6489 [astro-ph.CO].



\bibitem{Romano:2008rr} 
  A.~E.~Romano and M.~Sasaki,
  ``Effects of particle production during inflation,''
  Phys.\ Rev.\ D {\bf 78}, 103522 (2008)
  [arXiv:0809.5142 [gr-qc]].

\bibitem{Battefeld:2010rf} 
  D.~Battefeld, T.~Battefeld, H.~Firouzjahi and N.~Khosravi,
  ``Brane Annihilations during Inflation,''
  JCAP {\bf 1007}, 009 (2010)
  [arXiv:1004.1417 [hep-th]].
  
\bibitem{Firouzjahi:2010ga} 
  H.~Firouzjahi and S.~Khoeini-Moghaddam,
  ``Fields Annihilation and Particles Creation in DBI inflation,''
  JCAP {\bf 1102}, 012 (2011)
  [arXiv:1011.4500 [hep-th]].
  
\bibitem{Battefeld:2010vr} 
  D.~Battefeld, T.~Battefeld, J.~T.~Giblin, Jr. and E.~K.~Pease,
  ``Observable Signatures of Inflaton Decays,''
  JCAP {\bf 1102}, 024 (2011)
  [arXiv:1012.1372 [astro-ph.CO]].


\bibitem{Barnaby:2009dd} 
  N.~Barnaby and Z.~Huang,
  ``Particle Production During Inflation: Observational Constraints and Signatures,''
  Phys.\ Rev.\ D {\bf 80}, 126018 (2009)
  [arXiv:0909.0751 [astro-ph.CO]].

\bibitem{Barnaby:2010ke} 
  N.~Barnaby,
  ``On Features and Nongaussianity from Inflationary Particle Production,''
  Phys.\ Rev.\ D {\bf 82}, 106009 (2010)
  [arXiv:1006.4615 [astro-ph.CO]];
  N.~Barnaby,
  ``Nongaussianity from Particle Production During Inflation,''
  Adv.\ Astron.\  {\bf 2010}, 156180 (2010)
  [arXiv:1010.5507 [astro-ph.CO]].

\bibitem{Biswas:2010si} 
  T.~Biswas, A.~Mazumdar and A.~Shafieloo,
  ``Wiggles in the cosmic microwave background radiation: echoes from non-singular cyclic-inflation,''
  Phys.\ Rev.\ D {\bf 82}, 123517 (2010)
  [arXiv:1003.3206 [hep-th]].


\bibitem{Zarei:2008nr}
  M.~Zarei,
  ``Short Distance Physics and Initial State Effects on the CMB Power Spectrum
  and Cosmological Constant,''
  Phys.\ Rev.\  D {\bf 78}, 123502 (2008)
  [arXiv:0809.4312 [hep-th]].
  

\bibitem{Silverstein:2008sg} 
  E.~Silverstein and A.~Westphal,
  ``Monodromy in the CMB: Gravity Waves and String Inflation,''
  Phys.\ Rev.\ D {\bf 78}, 106003 (2008)
  [arXiv:0803.3085 [hep-th]].

\bibitem{Flauger:2009ab} 
  R.~Flauger, L.~McAllister, E.~Pajer, A.~Westphal and G.~Xu,
  ``Oscillations in the CMB from Axion Monodromy Inflation,''
  JCAP {\bf 1006}, 009 (2010)
  [arXiv:0907.2916 [hep-th]].

\bibitem{Flauger:2010ja} 
  R.~Flauger and E.~Pajer,
  ``Resonant Non-Gaussianity,''
  JCAP {\bf 1101}, 017 (2011)
  [arXiv:1002.0833 [hep-th]].

\bibitem{Bean:2008na} 
  R.~Bean, X.~Chen, G.~Hailu, S.~-H.~H.~Tye and J.~Xu,
  ``Duality Cascade in Brane Inflation,''
  JCAP {\bf 0803}, 026 (2008)
  [arXiv:0802.0491 [hep-th]].


\bibitem{White:2012ya} 
  J.~White, M.~Minamitsuji and M.~Sasaki,
  ``Curvature perturbation in multi-field inflation with non-minimal coupling,''
  arXiv:1205.0656 [astro-ph.CO].

\bibitem{Deruelle:2010ht} 
  N.~Deruelle and M.~Sasaki,
  ``Conformal equivalence in classical gravity: the example of 'veiled' General Relativity,''
  arXiv:1007.3563 [gr-qc].

\bibitem{Makino:1991sg} 
  N.~Makino and M.~Sasaki,
  ``The Density perturbation in the chaotic inflation with nonminimal coupling,''
  Prog.\ Theor.\ Phys.\  {\bf 86}, 103 (1991).

\bibitem{Tsujikawa:2004my} 
  S.~Tsujikawa and B.~Gumjudpai,
  ``Density perturbations in generalized Einstein scenarios and constraints on nonminimal couplings from the Cosmic Microwave Background,''
  Phys.\ Rev.\ D {\bf 69}, 123523 (2004)
  [astro-ph/0402185].

\bibitem{Koh:2005qp} 
  S.~Koh,
  ``Non-gaussianity in nonminimally coupled scalar field theory,''
  J.\ Korean Phys.\ Soc.\  {\bf 49}, S787 (2006)
  [astro-ph/0510030].

\bibitem{Koh:2010kg} 
  S.~Koh and M.~Minamitsuji,
  ``Non-minimally coupled hybrid inflation,''
  Phys.\ Rev.\ D {\bf 83}, 046009 (2011)
  [arXiv:1011.4655 [hep-th]].

\bibitem{Gong:2011qe} 
  J.~-O.~Gong, J.~-c.~Hwang, W.~-I.~Park, M.~Sasaki and Y.~-S.~Song,
  ``Conformal invariance of curvature perturbation,''
  JCAP {\bf 1109}, 023 (2011)
  [arXiv:1107.1840 [gr-qc]].

\bibitem{Kubota:2011re} 
  T.~Kubota, N.~Misumi, W.~Naylor and N.~Okuda,
  ``The Conformal Transformation in General Single Field Inflation with Non-Minimal Coupling,''
  JCAP {\bf 1202}, 034 (2012)
  [arXiv:1112.5233 [gr-qc]].


\bibitem{Kallosh:2010ug} 
  R.~Kallosh and A.~Linde,
  ``New models of chaotic inflation in supergravity,''
  JCAP {\bf 1011}, 011 (2010)
  [arXiv:1008.3375 [hep-th]].


\bibitem{Linde:2011nh} 
  A.~Linde, M.~Noorbala and A.~Westphal,
  ``Observational consequences of chaotic inflation with nonminimal coupling to gravity,''
  JCAP {\bf 1103}, 013 (2011)
  [arXiv:1101.2652 [hep-th]].

\bibitem{Ferrara:2010yw} 
  S.~Ferrara, R.~Kallosh, A.~Linde, A.~Marrani and A.~Van Proeyen,
  ``Jordan Frame Supergravity and Inflation in NMSSM,''
  Phys.\ Rev.\ D {\bf 82}, 045003 (2010)
  [arXiv:1004.0712 [hep-th]].


\bibitem{Ferrara:2010in} 
  S.~Ferrara, R.~Kallosh, A.~Linde, A.~Marrani and A.~Van Proeyen,
  ``Superconformal Symmetry, NMSSM, and Inflation,''
  Phys.\ Rev.\ D {\bf 83}, 025008 (2011)
  [arXiv:1008.2942 [hep-th]].

\bibitem{DeFelice:2011jm} 
  A.~De Felice, S.~Tsujikawa, J.~Elliston and R.~Tavakol,
  ``Chaotic inflation in modified gravitational theories,''
  JCAP {\bf 1108}, 021 (2011)
  [arXiv:1105.4685 [astro-ph.CO]].

\bibitem{Jamil:2009sq} 
  M.~Jamil, E.~N.~Saridakis and M.~R.~Setare,
  ``Holographic dark energy with varying gravitational constant,''
  Phys.\ Lett.\ B {\bf 679}, 172 (2009)
  [arXiv:0906.2847 [hep-th]].
  
  J.~Lu, E.~N.~Saridakis, M.~R.~Setare and L.~Xu,
  ``Observational constraints on holographic dark energy with varying gravitational constant,''
  JCAP {\bf 1003}, 031 (2010)
  [arXiv:0912.0923 [astro-ph.CO]].

\bibitem{Kofman:2007tr} 
  L.~Kofman and S.~Mukohyama,
  ``Rapid roll Inflation with Conformal Coupling,''
  Phys.\ Rev.\ D {\bf 77}, 043519 (2008)
  [arXiv:0709.1952 [hep-th]].

\bibitem{Bassett:2005xm} 
  B.~A.~Bassett, S.~Tsujikawa and D.~Wands,
  ``Inflation dynamics and reheating,''
  Rev.\ Mod.\ Phys.\  {\bf 78}, 537 (2006)
  [astro-ph/0507632].


\bibitem{Jain:2007yk} 
  B.~Jain and P.~Zhang,
  ``Observational Tests of Modified Gravity,''
  Phys.\ Rev.\ D {\bf 78}, 063503 (2008)
  [arXiv:0709.2375 [astro-ph]].

\bibitem{Afshordi:2008rd} 
  N.~Afshordi, G.~Geshnizjani and J.~Khoury,
  ``Do observations offer evidence for cosmological-scale extra dimensions?,''
  JCAP {\bf 0908}, 030 (2009)
  [arXiv:0812.2244 [astro-ph]].

\bibitem{Carroll:2006jn} 
  S.~M.~Carroll, I.~Sawicki, A.~Silvestri and M.~Trodden,
  ``Modified-Source Gravity and Cosmological Structure Formation,''
  New J.\ Phys.\  {\bf 8}, 323 (2006)
  [astro-ph/0607458].

\bibitem{Hu:2007pj} 
  W.~Hu and I.~Sawicki,
  ``A Parameterized Post-Friedmann Framework for Modified Gravity,''
  Phys.\ Rev.\ D {\bf 76}, 104043 (2007)
  [arXiv:0708.1190 [astro-ph]].






\bibitem{Weinberg:2008zzc} 
  S.~Weinberg,
  ``Cosmology,''
  Oxford, UK: Oxford Univ. Pr. (2008) 593 p
  
\bibitem{Wands:2000dp} 
  D.~Wands, K.~A.~Malik, D.~H.~Lyth and A.~R.~Liddle,
  ``A New approach to the evolution of cosmological perturbations on large scales,''
  Phys.\ Rev.\ D {\bf 62}, 043527 (2000)
  [astro-ph/0003278].

\bibitem{Lyth:2004gb} 
  D.~H.~Lyth, K.~A.~Malik and M.~Sasaki,
  ``A General proof of the conservation of the curvature perturbation,''
  JCAP {\bf 0505}, 004 (2005)
  [astro-ph/0411220].

  
\bibitem{Sasaki:1995aw}
  M.~Sasaki, E.~D.~Stewart,
  ``A General analytic formula for the spectral index of
 the density perturbations produced during inflation,''
  Prog.\ Theor.\ Phys.\  {\bf 95}, 71-78 (1996).
  [astro-ph/9507001].

\bibitem{Sasaki:1998ug} 
  M.~Sasaki and T.~Tanaka,
  ``Superhorizon scale dynamics of multiscalar inflation,''
  Prog.\ Theor.\ Phys.\  {\bf 99}, 763 (1998)
  [gr-qc/9801017].

\bibitem{Lyth-Liddle}
D. H. Lyth and A. R. Liddle,
 \textit{Primordial Density Perturbation} (Cambridge University Press, 2009).
 
\bibitem{Linde:1993cn}
  A.~D.~Linde,
  ``Hybrid inflation,''
  Phys.\ Rev.\  D {\bf 49}, 748 (1994)
  [arXiv:astro-ph/9307002].

\bibitem{Copeland:1994vg}
 A.~R.~Liddle, D.~H.~Lyth, E.~D.~Stewart and D.~Wands,
  ``False vacuum inflation with Einstein gravity,''
  Phys.\ Rev.\  D {\bf 49}, 6410 (1994)
  [arXiv:astro-ph/9401011].

 
\bibitem{Lyth:2010ch} 
  D.~H.~Lyth,
  ``Issues concerning the waterfall of hybrid inflation,''
  Prog.\ Theor.\ Phys.\ Suppl.\  {\bf 190}, 107 (2011)
  [arXiv:1005.2461 [astro-ph.CO]].

\bibitem{Abolhasani:2010kr}
  A.~A.~Abolhasani, H.~Firouzjahi,
  ``No Large Scale Curvature Perturbations during Waterfall of Hybrid Inflation,''
  Phys.\ Rev.\  {\bf D83}, 063513 (2011).
  [arXiv:1005.2934 [hep-th]].
  
\bibitem{Fonseca:2010nk}
  J.~Fonseca, M.~Sasaki, D.~Wands,
``Large-scale Perturbations from the Waterfall Field in Hybrid Inflation,''
  JCAP {\bf 1009}, 012 (2010).
  [arXiv:1005.4053 [astro-ph.CO]].

\bibitem{Abolhasani:2011yp} 
  A.~A.~Abolhasani, H.~Firouzjahi and M.~Sasaki,
  ``Curvature perturbation and waterfall dynamics in hybrid inflation,''
  JCAP {\bf 1110}, 015 (2011)
  [arXiv:1106.6315 [astro-ph.CO]].
  
\bibitem{Gong:2010zf}
  J.~-O.~Gong, M.~Sasaki,
  ``Waterfall field in hybrid inflation and curvature perturbation,''
  JCAP {\bf 1103}, 028 (2011).
  [arXiv:1010.3405 [astro-ph.CO]].

\bibitem{Lyth:2012yp} 
  D.~H.~Lyth,
  ``The hybrid inflation waterfall and the primordial curvature perturbation,''
  JCAP {\bf 1205}, 022 (2012)
  [arXiv:1201.4312 [astro-ph.CO]].

 
  
\bibitem{Clesse:2010iz} 
  S.~Clesse,
  ``Hybrid inflation along waterfall trajectories,''
  Phys.\ Rev.\ D {\bf 83}, 063518 (2011)
  [arXiv:1006.4522 [gr-qc]].

\bibitem{Martin:2011ib} 
  J.~Martin and V.~Vennin,
  ``Stochastic Effects in Hybrid Inflation,''
  Phys.\ Rev.\ D {\bf 85}, 043525 (2012)
  [arXiv:1110.2070 [astro-ph.CO]].

\bibitem{Mulryne:2011ni} 
  D.~Mulryne, S.~Orani and A.~Rajantie,
  ``Non-Gaussianity from the hybrid potential,''
  Phys.\ Rev.\ D {\bf 84}, 123527 (2011)
  [arXiv:1107.4739 [hep-th]].

\bibitem{Avgoustidis:2011em} 
  A.~Avgoustidis, S.~Cremonini, A.~-C.~Davis, R.~H.~Ribeiro, K.~Turzynski and S.~Watson,
  ``The Importance of Slow-roll Corrections During Multi-field Inflation,''
  JCAP {\bf 1202}, 038 (2012)
  [arXiv:1110.4081 [astro-ph.CO]].

\bibitem{Kodama:2011vs} 
  H.~Kodama, K.~Kohri and K.~Nakayama,
  ``On the waterfall behavior in hybrid inflation,''
  Prog.\ Theor.\ Phys.\  {\bf 126}, 331 (2011)
  [arXiv:1102.5612 [astro-ph.CO]].


\bibitem{Abolhasani:2010kn} 
  A.~A.~Abolhasani, H.~Firouzjahi and M.~H.~Namjoo,
  ``Curvature Perturbations and non-Gaussianities from Waterfall Phase Transition during Inflation,''
  Class.\ Quant.\ Grav.\  {\bf 28}, 075009 (2011)
  [arXiv:1010.6292 [astro-ph.CO]].
 
 
\bibitem{Bugaev:2011qt} 
  E.~Bugaev and P.~Klimai,
  ``Curvature perturbation spectra from waterfall transition, black hole constraints and non-Gaussianity,''
  JCAP {\bf 1111}, 028 (2011)
  [arXiv:1107.3754 [astro-ph.CO]].

\bibitem{Bugaev:2011wy} 
  E.~Bugaev and P.~Klimai,
  ``Formation of primordial black holes from non-Gaussian perturbations produced in a waterfall transition,''
  Phys.\ Rev.\ D {\bf 85}, 103504 (2012)
  [arXiv:1112.5601 [astro-ph.CO]].


\end{thebibliography}
\end{document}